\documentclass[12pt]{article}

\usepackage{times}
\usepackage{bbm}
\usepackage{float}
\usepackage{amsmath,amsfonts}
\usepackage{hyperref}
\usepackage{graphicx}
\usepackage{subfigure}
\usepackage{multirow}
\usepackage{blkarray}
\usepackage{tabularx}
\usepackage{adjustbox}
\usepackage{booktabs}
\usepackage{array}
\usepackage{amsthm,amscd,amsxtra,amsfonts,amsmath,amssymb,multirow}
\usepackage{algorithm,algorithmic}
\usepackage{epstopdf}
\usepackage{xcolor}
\usepackage{setspace}

\theoremstyle{definition}

\theoremstyle{remark}

\theoremstyle{example}

\theoremstyle{conjecture}

\newcommand {\R} {\mathbb R}

\title{Integration of persistent Laplacian and pre-trained transformer for protein solubility changes upon mutation}

\author
{JunJie Wee$^{1}$, Jiahui Chen$^{2}$, Kelin Xia$^{3,\dag}$ and Guo-Wei Wei$^{1,4,5,\ast}$\\
\normalsize{$^{1}$Department of Mathematics, Michigan State University, East Lansing, MI 48824, USA}\\
\normalsize{$^{2}$Department of Mathematical Sciences, University of Arkansas, Fayetteville, AR 72701, USA}\\
\normalsize{$^{3}$Division of Mathematical Sciences, School of Physical and Mathematical Sciences} \\
\normalsize{Nanyang Technological University, Singapore 637371}\\
\normalsize{$^{4}$Department of Biochemistry and Molecular Biology,}\\ 
\normalsize{Michigan State University, East Lansing, MI 48824, USA}\\
\normalsize{$^{5}$Department of Electrical and Computer Engineering,} \\
\normalsize{Michigan State University, East Lansing, MI 48824, USA}\\
\normalsize{$^\dag$ Address correspondences to Kelin Xia. E-mail: xiakelin@ntu.edu.sg}\\
\normalsize{$^\ast$ Address correspondences to Guo-Wei Wei. E-mail: weig@msu.edu}
}

\date{}

\begin{document}


\baselineskip24pt


\maketitle

\begin{abstract}
    Protein mutations can significantly influence protein solubility, which results in  altered  protein functions and leads to various diseases. Despite of tremendous effort, machine learning prediction of	protein solubility changes upon mutation remains a challenging task as indicated by the poor scores of normalized Correct Prediction Ratio (CPR). Part of the challenge stems from the fact that there is no three-dimensional (3D) structures for the wild-type and mutant proteins. This work integrates persistent Laplacians and pre-trained Transformer for the task. The Transformer, pretrained with hunderds of millions of protein sequences,  embeds wild-type and mutant sequences, while persistent Laplacians track the topological invariant change and homotopic shape evolution induced by mutations in 3D protein structures, which are rendered from AlphaFold2.  The resulting machine learning model was trained on an extensive data set labeled with three solubility types. Our model outperforms all existing predictive methods and improves the state-of-the-art up to  15\%.  

\end{abstract}

\newpage
\begin{spacing}{0.1}
\tableofcontents
\end{spacing}

\newpage

\section{Introduction}
Genetic mutations alter the genome sequence, leading to changes in the corresponding amino acid sequence of a protein. These alternations have far-reaching implications on the protein's structure, function, and stability, affecting attributes such as folding stability, binding affinity, and solubility. The consequences of protein mutations have been extensively studied in diverse fields such as evolutionary biology, cancer biology, immunology, directed evolution, and protein engineering~\cite{qiu2023persistent}.  Understanding the impact of genetic mutations on protein solubility is crucial in various fields, including protein engineering, drug discovery, and biotechnology. Accurately analyzing and predicting the impact of mutations on protein solubility is therefore crucial in many fields, facilitating the engineering of proteins with desirable functions. There are numerous intricately interconnected factors impacting protein solubility, ranging from amino acid sequence arrangement, post-translational modifications, protein-protein interactions,  to environmental conditions, such as solvent type, ion type and concentration, the presence of small molecules, temperature, etc. Unfortunately, the existing data set does not contain sufficient  information.  This complexity poses significant challenges for the accurate prediction and modeling of protein solubility, often requiring multifaceted computational approaches for reliable outcomes.

Computational predictions serve as a valuable complement to experimental mutagenesis analysis of  protein stability changes upon mutation. Such computational approaches offer several advantages, including being economical, efficient, and provide a viable alternative to labor-intensive site-directed mutagenesis experiments \cite{guerois2002predicting}. As a result, the development of accurate and reliable computational techniques for mutational impact prediction could substantially enhance the throughput and accessibility of research in protein engineering and drug discovery. 

Over the years, a variety of computational methods have been developed  to explore the effects of mutations on protein solubility, including but not limited to CamSol \cite{sormanni2015camsol}, OptSolMut \cite{tian2010scoring}, PON-Sol \cite{yang2016pon}, SODA \cite{paladin2017soda},   Solubis \cite{van2016solubis}, and others as summarized in a recent review \cite{vihinen2020solubility}. 
CamSol employs an algorithm to construct a residue-specific solubility profile, although no explicit method has been made publicly available. 
OptSolMut is trained on 137 samples, each featuring single or multiple mutations affecting solubility or aggregation. 
PON-Sol utilizes a random forest model trained on a dataset of 406 single amino acid substitutions labeled as solubility-increasing, solubility-decreasing, or exhibiting no change in solubility. 
SODA, which is based on the PON-Sol data, specifically targets samples with decreasing solubility \cite{paladin2017soda}. 
Solubis is an optimization tool that increases protein solubility and integrates interaction analysis from FoldX \cite{guerois2002predicting}, aggregation prediction from TANGO \cite{fernandez2004prediction}, and structural analysis from YASARA \cite{land2018yasara}. 
Recently, PON-Sol2 \cite{yang2021pon} extended the original PON-Sol dataset and employed a gradient boosting algorithm  for sequence-based predictions. 
Despite of intensive effort, the current prediction accuracy in terms of normalized Correct Prediction Ratio (CPR) remains very low, calling for innovative strategies.

Topological data analysis (TDA) is a relatively new approach for data science. Its main technique is  persistent homology \cite{edelsbrunner2022computational, zomorodian2004computing}. The essential idea of persistent homology is to construct a multiscale analysis of data in terms of topological invariants. The resulting changes of topological invariants over scales can be used to characterize the intricate structures  of data, leading to an unusually powerful approach in describing protein structure, flexibility, and folding \cite{xia2014persistent}.  
Persistent homology was integrated with machine learning for the classification of proteins in 2015 \cite{cang2015topological}, which was one the first integrations of TDA and machine learning,  and the predictions of mutation-induced protein stability changes  \cite{cang2017topologynet,cang2018integration} and protein-protein binding free energy changes \cite{wang2020topology,chen2020mutations}.  One of the major achievements of TDA is its winning of D3R Grand Challenges, an annual worldwide competition series in computer-aided drug design \cite{nguyen2019mathematical, nguyen2020mathdl}. A nearly comprehensive summary of the early success of TDA in biological science was given in a review \cite{nguyen2020review}.  

However, persistent homology only tracks the changes in topological invariants and cannot capture homotopic shape evolution of data over scales or induced by mutations. To overcome this limitation, Wei and coworkers introduced  persistent combinatorial Laplacians, also called persistent spectral graphs, on point clouds  \cite{wang2019persistent} and evolutionary de Rham-Hodge method on manifolds \cite{chen2019evolutionary} in 2019. The essence of these methods is the persistent topological Laplacians (PTLs) either on point clouds or on manifolds. PTLs not only fully capture the topological invariants in its harmonic spectra as those given by persistent homology, but also capture the homotopic shape evolution of data during the multiscale analysis or a mutation process. PTLs were applied to the predictions of protein flexibility \cite{wang2020persistent} and protein-ligand binding free energies \cite{meng2021persistent}, protein–protein interactions\cite{wee2022persistent, bi2023multiscale}, and protein engineering \cite{qiu2023persistent}. The most remarkable accomplishment by persistent Laplacian is its accurate forecasting of emerging dominant SARS-CoV-2 variants BA.4 and BA.5 about two months in advance \cite{chen2022persistent}.


However, TDA approaches depend on the biomolcular structures, which may not be available. In fact, many proteins involved in the present study do not have 3D structures. In recent years, advanced natural language processing (NLP) models, including Transformers and long short-term memory (LSTM), have been widely implemented across various domains to  extract  protein sequence information. For example,  Tasks Assessing Protein Embedding  (TAPE)  introduced three different architectures, namely transformer, dilated residual network (ResNet), and LSTM  \cite{rao2019evaluating}. 
Additionally, LSTM-based models like Bepler \cite{bepler2018learning} and UniRep \cite{alley2019unified} have been developed. Additionally,  large-scale protein transformer models trained on extensive datasets comprising hundreds of millions of sequences have made significant advancements in the field.   
These models, including Evolutionary Scale Modeling (ESM) \cite{rives2021biological} and ProtTrans \cite{vaswani2017attention,Devlin2019BERTPO}, have exhibited exceptional performance in capturing a variety of protein properties. ESM, for instance, allows for fine-tuning based on either downstream task data or local multiple sequence alignments \cite{NEURIPS2021_f51338d7}. In the present work, we leverage the pre-trained ESM transformer model to extract crucial information from protein sequences.

In this work, we will integrate transformer-based sequence embedding and persistent topological  Laplacians to predict protein solubility changes upon mutation. While sequence-based models can be applied without 3D structural information, 
the PTL-based features require high-quality structures. We generate  3D structures of wild type proteins from AlphaFold2 \cite{jumper2021highly} to facilitate topological embedding. By combining both embedding  approaches, they naturally complement each other in classifying protein solubility changes upon mutation. These embeddings are fed into an ensemble classifier, gradient boosted trees (GBT), to build a machine learning model, called TopLapGBT. We validate TopLapGBT on the classification of  protein solubility changes upon mutation. We demonstrate that this integrated machine learning model gives rise to a substantial improvement as compared to existing state-of-the-art models. Residue-Similarity plots are also applied to assess how well the TopLapGBT model classify  three solubility labels. 

\section{Results}

\subsection{Overview of TopLapGBT}
\begin{figure}[!ht]
	\centering
	\includegraphics[width=\textwidth]{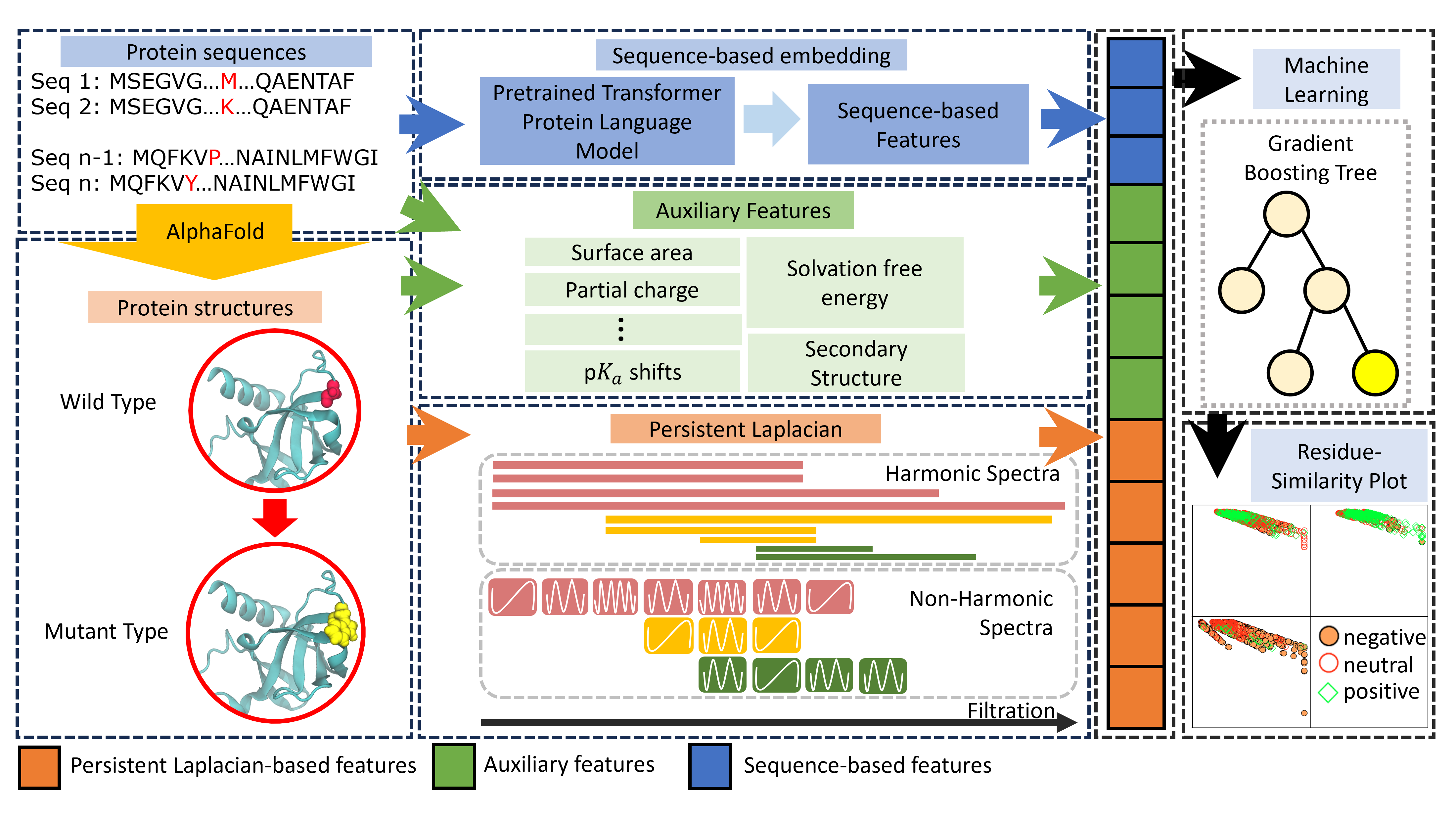}
	\caption{The illustration of the workflow for TopLapGBT. Protein sequences are first preprocessed by AlphaFold 2 to generate wild type protein structures. Mutant proteins are generated from the Jackal software \cite{xiang2002jackal}. The structure-based features from persistent Laplacian, auxiliary and sequence-based features are then concatenated to form a long feature input for gradient boosting tree to classify the protein solubility changes upon mutation. The predicted labels are also analyzed on Residue-Similarity (R-S) plots.}
	\label{fig:model}
\end{figure}

TopLapGBT integrates both structure-based and sequence-based features, derived from protein structures and sequences respectively, into a unified model. Our architecture comprises three distinct embedding modules: persistent Laplacian-based embeddings, sequence-based embeddings, and auxiliary feature embeddings, all of which feed into an ensemble classifier as depicted in Figure \ref{fig:model}.

In the persistent Laplacian-based feature embedding module, we employ persistent Laplacian techniques to generate features that encapsulate the structural attributes of proteins both pre- and post-mutation. This approach is particularly effective in capturing the structural alterations induced by mutations within the localized neighborhoods of the mutation sites. Mathematically, the persistent Laplacian builds a sequence of simplicial complexes through a filtration process, thereby characterizing atom-atom interactions across multiple scales (details in the Methods section).
In the sequence-based feature embedding module, a pre-trained transformer model generates latent feature vectors extracted from protein sequences. Specifically, the transformer model used here is a 650M-parameter protein language model, trained on a corpus of 250M protein sequences spanning multiple organisms \cite{rao2021msa}.
Finally, the auxiliary feature embedding module incorporates a variety of attributes such as surface area, partial charge, p$K_a$ shifts, solvation free energy, and secondary structural information, synthesized from both protein sequences and structures. 
These three distinct sets of feature embeddings are subsequently concatenated to produce a comprehensive feature vector. This vector is then fed into a gradient-boosting tree classifier to categorize the mutation-induced samples.

\subsection{Performance of TopLapGBT on PON-Sol2 dataset}
In our study, we utilize the dataset employed by PON-Sol2 as detailed in \cite{yang2021pon}. The dataset is comprised of 6,328 mutation samples, originating from 77 distinct proteins. These samples are categorized into three labels: decrease in solubility, increase in solubility, and no change in solubility. Specifically, the dataset contains 3,136 samples demonstrating a decrease in solubility, 1,026 samples showing an increase, and 2,166 samples with no change. Notably, the dataset exhibits a class imbalance, with a ratio of $1:0.69:0.34$, indicating a bias towards samples that exhibit a decrease in solubility.
To assess the performance of our model, we initially carry out a random 10-fold cross-validation on the dataset. Subsequently, an independent blind test prediction is executed to provide further validation of the model's efficacy.

In Table \ref{tab:res1}, we present a comparative analysis of the performance of existing classifiers by PON-Sol \cite{yang2016pon} and PON-Sol2 \cite{yang2021pon} against our proposed model, TopLapGBT, using 10-fold cross-validation. It should be noted that PON-Sol2 incorporates feature selection techniques such as recursive feature elimination (RFE). To provide a robust assessment of TopLapGBT's performance, we conduct 10 repeated runs, and the mean values of these runs are reported to account for any randomness in the model's output.

\begin{table}[!ht]
	\caption{Comparison of performance metrics between TopLapGBT and both single layer and double layer classifiers of PON-Sol2 in the 10-fold crossvalidation. The negative solubility samples are denoted as "-" whereas the positive solubility change samples are denoted as "+". The samples with no solubility change are denoted as "N". Performance metrics include the positive predicted values (PPV), negative predicted values (NPV), sensitivity, specificity, correct prediction ratio (CPR) and generalised correlation (GC$^2$). PPV refers to the proportions of positive predictions for each solubility class while NPV refers to the proportions of negative predictions for each solubility class. CPR calculates the percentage of correctly classified samples while GC$^2$ measures the correlation coefficient of the classification. All normalized metrics are also reported. For each metric, the first value is without normalization while the second one is with normalization.}
	\label{tab:res1}
	\centering
	\resizebox{\textwidth}{!}{
		\begin{tabular}{|cc|cccccc|}
			\hline
			\multicolumn{2}{|c|}{\multirow{5}{*}{\begin{tabular}[c]{@{}c@{}}Performance\\ Metric\end{tabular}}} & \multicolumn{6}{c|}{Model}
		    \\ \cline{3-8} 
			\multicolumn{2}{|c|}{}                                                                              & \multicolumn{4}{c|}{PON-Sol2 \cite{yang2021pon}}                                                                                                                                                                                                                                                                                                                                                                                             &\multicolumn{1}{c|}{TopGBT} 
			& TopLapGBT                                                                          \\ \cline{3-8} 
			\multicolumn{2}{|c|}{}                                                                              & \multicolumn{2}{c|}{Single Three-Class Classifier}                                                                                                                                                          & \multicolumn{2}{c|}{Two-Layer Three-Class Classifier}                                                                                                                                                       & \multicolumn{1}{c|}{-}         & -                                                                       \\ \cline{3-8} 
			\multicolumn{2}{|c|}{}                                                                              & \multicolumn{1}{c|}{All Features}                                                                    & \multicolumn{1}{c|}{\begin{tabular}[c]{@{}c@{}}30 Features \\ Selected by RFE\end{tabular}}          & \multicolumn{1}{c|}{All Features}                                                                    & \multicolumn{1}{c|}{\begin{tabular}[c]{@{}c@{}}34 Features\\ Selected by RFE\end{tabular}}           & \multicolumn{1}{c|}{-}     & -                                                                          \\ \hline
			\multicolumn{1}{|c|}{PPV}                & \begin{tabular}[c]{@{}c@{}}-\\ N\\ +\end{tabular}        & \multicolumn{1}{c|}{\begin{tabular}[c]{@{}c@{}}0.842/0.742\\ 0.657/0.536\\ 0.563/0.730\end{tabular}} & \multicolumn{1}{c|}{\begin{tabular}[c]{@{}c@{}}0.835/0.729\\ 0.658/0.543\\ 0.586/0.752\end{tabular}} & \multicolumn{1}{c|}{\begin{tabular}[c]{@{}c@{}}0.875/0.793\\ 0.635/0.521\\ 0.520/0.696\end{tabular}} & \multicolumn{1}{c|}{\begin{tabular}[c]{@{}c@{}}0.869/0.781\\ 0.647/0.534\\ 0.538/0.714\end{tabular}}
			& \multicolumn{1}{c|}{\begin{tabular}[c]{@{}c@{}}0.868/0.785\\ 0.686/0.554\\ 0.646/0.797\end{tabular}}
			& \begin{tabular}[c]{@{}c@{}}0.873/0.797\\ 0.681/0.557\\ 0.627/0.779\end{tabular} \\ \hline
			\multicolumn{1}{|c|}{NPV}                & \begin{tabular}[c]{@{}c@{}}-\\ N\\ +\end{tabular}        & \multicolumn{1}{c|}{\begin{tabular}[c]{@{}c@{}}0.913/0.954\\ 0.841/0.824\\ 0.877/0.737\end{tabular}} & \multicolumn{1}{c|}{\begin{tabular}[c]{@{}c@{}}0.901/0.947\\ 0.847/0.832\\ 0.877/0.738\end{tabular}} & \multicolumn{1}{c|}{\begin{tabular}[c]{@{}c@{}}0.893/0.942\\ 0.847/0.829\\ 0.877/0.736\end{tabular}} & \multicolumn{1}{c|}{\begin{tabular}[c]{@{}c@{}}0.891/0.941\\ 0.855/0.838\\ 0.878/0.739\end{tabular}}
			& \multicolumn{1}{c|}{\begin{tabular}[c]{@{}c@{}}0.932/0.965\\ 0.864/0.849\\ 0.886/0.749\end{tabular}}
			& \begin{tabular}[c]{@{}c@{}}0.931/0.964\\ 0.858/0.842\\ 0.888/0.757\end{tabular} \\ \hline
			\multicolumn{1}{|c|}{Sensitivity}        & \begin{tabular}[c]{@{}c@{}}-\\ N\\ +\end{tabular}        & \multicolumn{1}{c|}{\begin{tabular}[c]{@{}c@{}}0.919/0.919\\ 0.701/0.701\\ 0.329/0.329\end{tabular}} & \multicolumn{1}{c|}{\begin{tabular}[c]{@{}c@{}}0.906/0.906\\ 0.717/0.717\\ 0.326/0.326\end{tabular}} & \multicolumn{1}{c|}{\begin{tabular}[c]{@{}c@{}}0.892/0.892\\ 0.724/0.724\\ 0.336/0.336\end{tabular}} & \multicolumn{1}{c|}{\begin{tabular}[c]{@{}c@{}}0.891/0.891\\ 0.738/0.738\\ 0.340/0.340\end{tabular}}
			& \multicolumn{1}{c|}{\begin{tabular}[c]{@{}c@{}}0.937/0.937\\ 0.752/0.752\\ 0.359/0.359\end{tabular}}
			& \begin{tabular}[c]{@{}c@{}}0.934/0.934\\ 0.735/0.735\\ 0.395/0.395\end{tabular} \\ \hline
			\multicolumn{1}{|c|}{Specificity}        & \begin{tabular}[c]{@{}c@{}}-\\ N\\ +\end{tabular}        & \multicolumn{1}{c|}{\begin{tabular}[c]{@{}c@{}}0.831/0.839\\ 0.812/0.697\\ 0.948/0.938\end{tabular}} & \multicolumn{1}{c|}{\begin{tabular}[c]{@{}c@{}}0.825/0.831\\ 0.807/0.697\\ 0.954/0.947\end{tabular}} & \multicolumn{1}{c|}{\begin{tabular}[c]{@{}c@{}}0.875/0.883\\ 0.785/0.667\\ 0.938/0.927\end{tabular}} & \multicolumn{1}{c|}{\begin{tabular}[c]{@{}c@{}}0.868/0.874\\ 0.792/0.678\\ 0.941/0.932\end{tabular}}
			& \multicolumn{1}{c|}{\begin{tabular}[c]{@{}c@{}}0.860/0.872\\ 0.821/0.697\\ 0.962/0.954\end{tabular}}
			& \begin{tabular}[c]{@{}c@{}}0.867/0.881\\ 0.823/0.707\\ 0.953/0.944\end{tabular} \\ \hline
			\multicolumn{2}{|c|}{CPR}                                                                           & \multicolumn{1}{c|}{0.747/0.650}                                                                     & \multicolumn{1}{c|}{0.746/0.650}                                                                     & \multicolumn{1}{c|}{0.743/0.651}                                                                     & \multicolumn{1}{c|}{0.747/0.656}    
			& \multicolumn{1}{c|}{0.780/0.682}                                                                  & \textbf{0.792/0.688}                                                            \\ \hline
			\multicolumn{2}{|c|}{GC$^2$}                                                                           & \multicolumn{1}{c|}{0.317/0.298}                                                                     & \multicolumn{1}{c|}{0.309/0.289}                                                                     & \multicolumn{1}{c|}{0.322/0.313}                                                                     & \multicolumn{1}{c|}{0.323/0.312}                  
			& \multicolumn{1}{c|}{0.371/0.354}                                     & \textbf{0.376/0.361}                                                            \\ \hline
	\end{tabular}}
\end{table}

Performance evaluation of our model, TopLapGBT, is conducted using a range of metrics, including Positive Predictive Value (PPV), Negative Predictive Value (NPV), Sensitivity, Specificity, Correct Prediction Ratio (CPR), and Generalized Squared Correlation (GC$^2$). PPV and NPV quantify the proportions of correct positive and negative predictions for each solubility class, respectively. Given that we are dealing with a $K$-class problem with three distinct solubility classes, CPR and GC$^2$ are particularly relevant for providing a holistic view of the model's performance \cite{baldi2000assessing}. Specifically, CPR measures the overall accuracy of the model, while GC$^2$ quantifies the correlation coefficient of the classification, ranging from 0 to 1. Larger values for these metrics denote better performance. Importantly, due to the class imbalance in the number of mutation samples across the categories, all performance metrics are normalized to ensure a robust and reliable evaluation of the model's efficacy (further details are elaborated in the Methods section).

The proposed model, TopLapGBT, demonstrates significant performance gains over existing featurization methods in PON-Sol2 across all evaluation metrics \cite{yang2021pon}. Specifically, normalized CPR and GC$^2$ scores of TopLapGBT stand at 0.688 and 0.361, marking improvements of 4.88\% and 15.71\% over PON-Sol2, respectively. These gains underscore the merit of incorporating both structure-based and sequence-based features into the model. 
To elucidate the contribution of Persistent Laplacian (PL)-based features, we also present a comparative analysis with our TopGBT model in Table \ref{tab:res1}. The TopGBT model utilizes persistent homology-based embeddings alongside auxiliary and pre-trained transformer features. While TopGBT still outperforms all existing PON-Sol2 models, the incorporation of PL-based features in TopLapGBT leads to an incremental improvement of 1\% and 2\% in CPR and GC$^2$ metrics, respectively. This validates our approach of leveraging Persistent Laplacian to comprehensively capture both the topological and homotopic nuances in the evolution of protein structures.

\subsection{Performance of TopLapGBT on independent test set}

\begin{table}[!ht]
	\caption{Performance of TopLapGBT with existing state-of-the-art models on the independent blind test classification. The negative solubility samples are denoted as "-" whereas the positive solubility change samples are denoted as "+". The samples with no solubility change are denoted as "N". Performance metrics include the positive predicted values (PPV), negative predicted values (NPV), sensitivity, specificity, correct prediction ratio (CPR) and generalised correlation (GC$^2$). PPV refers to the proportions of positive predictions for each solubility class while NPV refers to the proportions of negative predictions for each solubility class. CPR calculates the percentage of correctly classified samples while GC$^2$ measures the correlation coefficient of the classification. All normalized metrics are also reported. For each metric, the first value is without normalization while the second one is with normalization.}
	\label{tab:res2}
	\centering
	\resizebox{\textwidth}{!}{ 
		\begin{tabular}{|cc|lllllccc|}
			\hline
			\multicolumn{2}{|c|}{\multirow{2}{*}{\begin{tabular}[c]{@{}c@{}}Performance \\ Metric\end{tabular}}} & \multicolumn{8}{c|}{Independent Test}                                                                                                                                                                                                                                                                                                                                                                                                                                                                                                                                                                                                                                                                                 \\ \cline{3-10} 
			\multicolumn{2}{|c|}{}                                                                               & \multicolumn{1}{c|}{PON-Sol \cite{yang2016pon}}                                                                         &
			\multicolumn{1}{c|}{\begin{tabular}[c]{@{}c@{}}SODA\end{tabular}} &
			 \multicolumn{1}{c|}{\begin{tabular}[c]{@{}c@{}}SODA(5 as \\ Threshold)\end{tabular}}                 & \multicolumn{1}{c|}{\begin{tabular}[c]{@{}c@{}}SODA(10 as \\ Threshold)\end{tabular}}                & \multicolumn{1}{c|}{\begin{tabular}[c]{@{}c@{}}SODA(17 as \\ Threshold)\end{tabular}}                & \multicolumn{1}{c|}{PON-Sol2 \cite{yang2021pon}}                                                                        & \multicolumn{1}{c|}{TopGBT}          & \multicolumn{1}{c|}{TopLapGBT}                                                                                                                                       \\ \hline
			\multicolumn{1}{|c|}{PPV}                 & \begin{tabular}[c]{@{}c@{}}-\\ N\\ +\end{tabular}        & \multicolumn{1}{l|}{\begin{tabular}[c]{@{}l@{}}0.593/0.428\\ 0.427/0.385\\ 0.151/0.373\end{tabular}} &
			\multicolumn{1}{l|}{\begin{tabular}[c]{@{}l@{}}0.427/0.258\\ 
			NaN/NaN\\ 0.080/0.229\end{tabular}} &
			\multicolumn{1}{l|}{\begin{tabular}[c]{@{}l@{}}0.606/0.428\\ 0.425/0.365\\ 0.047/0.149\end{tabular}} & \multicolumn{1}{l|}{\begin{tabular}[c]{@{}l@{}}0.673/0.468\\ 0.397/0.357\\ 0.060/0.184\end{tabular}} & \multicolumn{1}{l|}{\begin{tabular}[c]{@{}l@{}}0.742/0.585\\ 0.383/0.350\\ 0.098/0.284\end{tabular}} & \multicolumn{1}{c|}{\begin{tabular}[c]{@{}c@{}}0.804/0.643\\ 0.600/0.475\\ 0.233/0.472\end{tabular}} &
			\multicolumn{1}{c|}{\begin{tabular}[c]{@{}c@{}}0.781/0.649\\ 0.617/0.462\\ 0.524/0.761\end{tabular}} & \multicolumn{1}{c|}{\begin{tabular}[c]{@{}c@{}}0.789/0.645\\ 0.624/0.475\\ 0.476/0.718\end{tabular}}  \\ \hline
			\multicolumn{1}{|c|}{NPV}                 & \begin{tabular}[c]{@{}c@{}}-\\ N\\ +\end{tabular}        & \multicolumn{1}{l|}{\begin{tabular}[c]{@{}l@{}}0.514/0.691\\ 0.685/0.700\\ 0.881/0.693\end{tabular}} &
			\multicolumn{1}{l|}{\begin{tabular}[c]{@{}l@{}}0.373/0.537\\ 
			0.642/0.667\\ 0.832/0.605\end{tabular}} &
			\multicolumn{1}{l|}{\begin{tabular}[c]{@{}l@{}}0.508/0.684\\ 0.761/0.739\\ 0.848/0.633\end{tabular}} & \multicolumn{1}{l|}{\begin{tabular}[c]{@{}l@{}}0.502/0.677\\ 0.797/0.782\\ 0.858/0.649\end{tabular}} & \multicolumn{1}{l|}{\begin{tabular}[c]{@{}l@{}}0.501/0.677\\ 0.797/0.782\\ 0.858/0.649\end{tabular}} & \multicolumn{1}{c|}{\begin{tabular}[c]{@{}c@{}}0.794/0.887\\ 0.804/0.793\\ 0.879/0.684\end{tabular}} &
			\multicolumn{1}{c|}{\begin{tabular}[c]{@{}c@{}}0.843/0.920\\ 0.816/0.795\\ 0.881/0.692\end{tabular}} & \multicolumn{1}{c|}{\begin{tabular}[c]{@{}c@{}}0.842/0.918\\ 0.826/0.809\\ 0.880/0.688\end{tabular}}  \\ \hline
			\multicolumn{1}{|c|}{Sensitivity}         & \begin{tabular}[c]{@{}c@{}}-\\ N\\ +\end{tabular}        & \multicolumn{1}{l|}{\begin{tabular}[c]{@{}l@{}}0.263/0.263\\ 0.456/0.456\\ 0.448/0.448\end{tabular}} &
			\multicolumn{1}{l|}{\begin{tabular}[c]{@{}l@{}}0.488/0.488\\ 0.000/0.000\\ 0.253/0.253\end{tabular}} &
			\multicolumn{1}{l|}{\begin{tabular}[c]{@{}l@{}}0.195/0.195\\ 0.759/0.759\\ 0.069/0.069\end{tabular}} & \multicolumn{1}{l|}{\begin{tabular}[c]{@{}l@{}}0.098/0.098\\ 0.886/0.886\\ 0.057/0.057\end{tabular}} & \multicolumn{1}{l|}{\begin{tabular}[c]{@{}l@{}}0.068/0.068\\ 0.954/0.954\\ 0.046/0.046\end{tabular}} & \multicolumn{1}{c|}{\begin{tabular}[c]{@{}c@{}}0.802/0.802\\ 0.671/0.671\\ 0.161/0.161\end{tabular}} &
			\multicolumn{1}{c|}{\begin{tabular}[c]{@{}c@{}}0.867/0.867\\ 0.692/0.692\\ 0.126/0.126\end{tabular}} & \multicolumn{1}{c|}{\begin{tabular}[c]{@{}c@{}}0.864/0.864\\ 0.713/0.713\\ 0.115/0.115\end{tabular}} \\ \hline
			\multicolumn{1}{|c|}{Specificity}         & \begin{tabular}[c]{@{}c@{}}-\\ N\\ +\end{tabular}        & \multicolumn{1}{l|}{\begin{tabular}[c]{@{}l@{}}0.812/0.824\\ 0.659/0.636\\ 0.617/0.623\end{tabular}} &
			\multicolumn{1}{l|}{\begin{tabular}[c]{@{}l@{}}0.318/0.297\\ 
			1.000/1.000\\ 0.558/0.573\end{tabular}} &
			\multicolumn{1}{l|}{\begin{tabular}[c]{@{}l@{}}0.867/0.869\\ 0.426/0.340\\ 0.786/0.802\end{tabular}} & \multicolumn{1}{l|}{\begin{tabular}[c]{@{}l@{}}0.951/0.944\\ 0.249/0.204\\ 0.863/0.872\end{tabular}} & \multicolumn{1}{l|}{\begin{tabular}[c]{@{}l@{}}0.975/0.976\\ 0.144/0.116\\ 0.936/0.942\end{tabular}} & \multicolumn{1}{c|}{\begin{tabular}[c]{@{}c@{}}0.796/0.777\\ 0.751/0.630\\ 0.920/0.910\end{tabular}} & \multicolumn{1}{c|}{\begin{tabular}[c]{@{}c@{}}0.747/0.765\\ 0.760/0.597\\ 0.983/0.980\end{tabular}}
			& \multicolumn{1}{c|}{\begin{tabular}[c]{@{}c@{}}0.759/0.763\\ 0.760/0.606\\ 0.981/0.977\end{tabular}}  \\ \hline
			\multicolumn{2}{|c|}{CPR}                                                                            & \multicolumn{1}{l|}{0.356/0.389}                                                                     &
			\multicolumn{1}{l|}{0.282/0.247}                                                                     &
			\multicolumn{1}{l|}{0.381/0.341}                                                                     & \multicolumn{1}{l|}{0.375/0.347}                                                                     & \multicolumn{1}{l|}{0.382/0.356}                                                                     & \multicolumn{1}{c|}{0.671/0.545}                                                                     & \multicolumn{1}{c|}{0.707/0.562}                                                                     &
			\multicolumn{1}{c|}{\textbf{0.711/0.564}  }                                                                                                                           \\ \hline
			\multicolumn{2}{|c|}{GC$^2$}                                                                            & \multicolumn{1}{l|}{0.010/0.011}                                                                     &
			\multicolumn{1}{l|}{NaN/NaN}                                                                     &
			\multicolumn{1}{l|}{0.041/0.045}                                                                     & \multicolumn{1}{l|}{0.022/0.022}                                                                     & \multicolumn{1}{l|}{0.016/0.016}                                                                     & \multicolumn{1}{c|}{0.181/0.157}                                                                     & \multicolumn{1}{c|}{0.205/0.184}                                                                     &
			\multicolumn{1}{c|}{\textbf{0.206/0.185}  }                                                                                                                            \\ \hline
	\end{tabular}}
\end{table}

To robustly assess the performance of TopLapGBT, we subjected it to an independent test using the same dataset employed by PON-Sol2 \cite{yang2021pon}. In this validation, TopLapGBT consistently outperformed all five existing models, as evidenced in Table \ref{tab:res2}. Specifically, TopLapGBT registers a normalized CPR of 0.564 and a normalized GC$^2$ of 0.185, surpassing PON-Sol2 by 3.49\% and 17.83\%, respectively. Relative to TopGBT, the inclusion of PL-based features in TopLapGBT yielded incremental gains in both CPR and GC$^2$ metrics, thereby further substantiating the utility of Persistent Laplacian in capturing the homotopic shape evolution within protein structures.

\section{Discussion}

The performance of machine learning models generally relies on the nature of the input features. In our model, the PL-based features depend on one main element which is the quality of the protein structures from AlphaFold 2 (AF2). The quality of AF2 structures are crucial in determining the performance of TopLapGBT. Recently, AF2 structures have been reported to achieve comparable performance to nuclear magnetic resonance (NMR) structures while ensemble methods can be used to enhance the performance by combining multiple NMR structures \cite{qiu2023persistent}. This allows AF2 structures to serve as a practical substitute for experimental structural data. Although AF2 structures are not as reliable as X-ray structures, the fusion of sequence-based pre-trained transformer features and PL-based features provides robust featurization even for low quality AF2 structural data. PL elucidates the precise mutation geometry and topology, while sequence-based pre-trained transformer features capture evolutionary patterns from an extensive sequence library. This synergy holds significance and can be applied to a diverse range of other challenges in the field of biomolecular research. For the rest of this section, we analyze the model's performance based on the region of the mutations and the type of mutations. We also discuss the performance of different feature types using the Residue-Similarity plots.

\subsection{Performance analysis based on different mutation regions}
\begin{figure}[!ht]
	\centering
	\includegraphics[width=.9\textwidth]{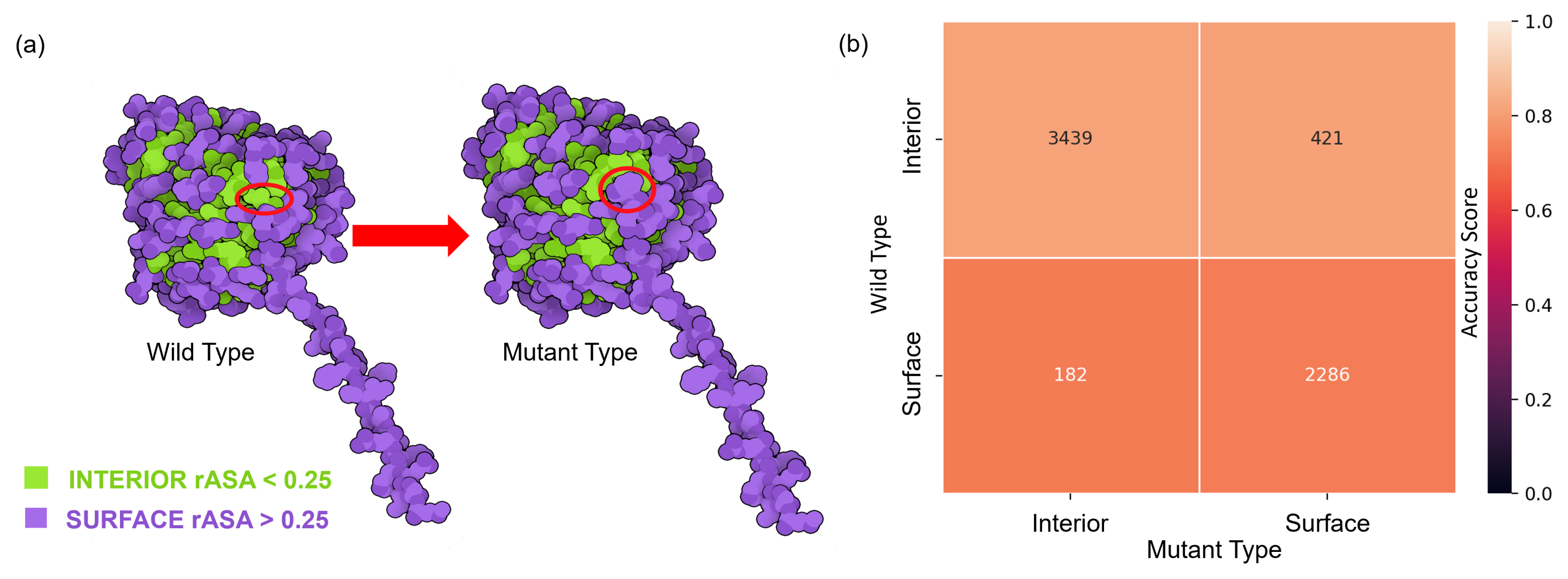}
	\caption{(a) The definitions of the structural regions on the protein label 213133708 with mutation ID: I283W. For both wild type and mutant type, amino acids in the proteins are classified under surface or interior regions based on the rASA of the residue. The residue ID 283 of protein label 213133708 was mutated from isoleucine (interior region) to trytophan (surface region). Structures are plotted with the software Illustrate\cite{goodsell2019illustrate}. 
	(b) A comparison of performance of TopLapGBT among different mutation region types. The $y$-axis represents the region type for the original residue and the $x$-axis represents the region type for the mutated residue. The numbers indicated in each cell corresponds to the number of mutation samples in each region-region mutation pair. The accuracy scores (CPR) for both interior-interior and interior-surface are 0.813 and 0.812 while the accuracy score for both surface-interior and surface-surface are 0.725 and 0.730.}
	\label{fig:region}
\end{figure}

To delve deeper into the model's performance, we categorize mutation samples based on their structural regions: interior and surface, as depicted in Figure \ref{fig:region} pre- and post-mutations. These regions are defined by their relative accessible solvent area (rASA), using a cutoff value $c$. A residue at the mutation site is classified as buried or interior if its rASA falls below this cutoff. While the discrete nature of $c$ initially raised concerns, given that amino acids have a continuous exposure profile, empirical analyses on databases from organisms like Escherichia coli, Saccharomyces cerevisiae, and Homo sapiens have shown that an optimal rASA cutoff of approximately 25\% is effective for distinguishing between surface and interior residues \cite{levy2010simple}. In our analysis, we apply this framework to identify surface and interior residues in the solubility dataset. We observe that some mutation sites undergo a regional transition, moving from one structural domain to another, consequent to the mutation.


To gain nuanced insights into TopLapGBT's performance, we segment the results according to the mutation's structural location within the protein. We present these segmentations as heatmap plots that delineate both mutation regions and amino acid types. Structural regions are defined based on relative accessible surface area (rASA) \cite{levy2010simple}. By categorizing residues as either interior or surface, we can examine the influence of continuous amino acid exposure on solubility change classification post-mutation. Figure \ref{fig:region}(b) displays accuracy scores for four types of mutations: interior-interior, interior-surface, surface-interior, and surface-surface. TopLapGBT attains an average accuracy score of 0.770 across these categories. Extended data in Figure S1 further breaks down accuracy scores for all 20 distinct amino acids within each region-pair, revealing variations in residue-residue pair performance.

\subsection{Performance analysis based on different mutation types}
\begin{figure}[!ht]
	\centering
	\includegraphics[width=1\textwidth]{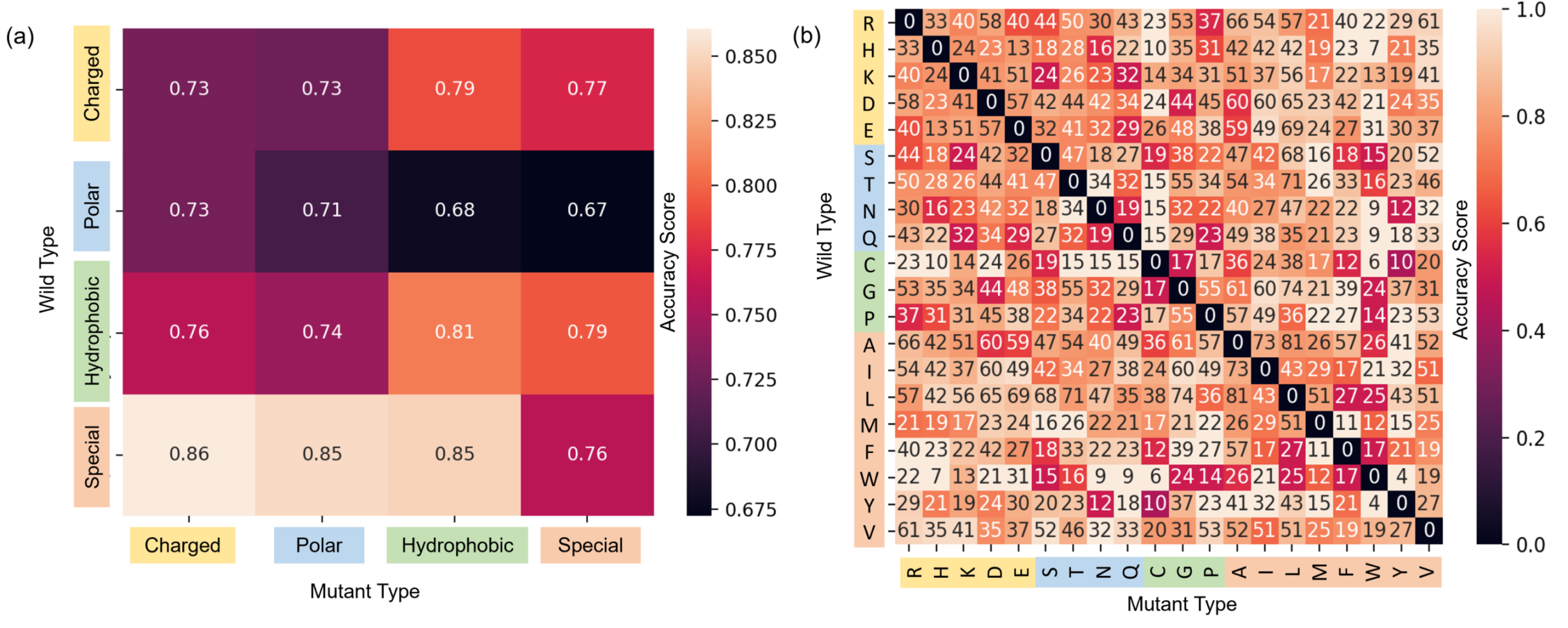}
	\caption{A comparison of 10-fold cross validation accuracy scores (CPR) for (a) different mutation groups and (b) its associated amino acid types. The $y$-axis labels the residue type of the original protein, whereas the $x$-axis labels the residue type of the mutant. The squares colored in black in (b) have zero mutation samples. For a reverse mutation, the labels are taken with reverse solubility change unless the change is zero.}
	\label{fig:amino}
\end{figure}

Switching focus to mutation types, our model's capability in classifying solubility changes also merits exploration across the 20 distinct amino acid types in the dataset. In addition to this, we subgroup amino acids as charged, polar, hydrophobic, or special case. Table S1 enumerates the sample counts for each mutation group pair. Figure \ref{fig:amino}(a) displays accuracy scores for each mutation group pair, while Figure \ref{fig:amino}(b) shows scores for each amino acid pair. Notably, the special-charged and special-polar groups register the highest accuracy, whereas the polar-hydrophobic and polar-special groups underperform. One plausible reason could be the inherent complexity in accurately classifying mutations with non-negative solubility changes. It's worth noting that PON-Sol2 employed a two-layer classifier to improve classification \cite{yang2021pon}. Our results indicate that TopLapGBT surpasses the performance of this two-layer system.

\subsection{Feature analysis based on Residue-Similarity plots}
The Residue Similarity Index (RSI) serves as a potent metric for evaluating the efficacy of dimensionality reduction in both clustering and classification contexts \cite{hozumi2023preprocessing}. RSI has proven its value in generating classification accuracy scores that align well with supervised methods in single-cell typing. When applied to our solubility change dataset, Residue-Similarity (R-S) plots can be constructed to scrutinize how the Residue Index (RI) and Similarity Index (SI) may indicate the quality of cluster separation.

\begin{figure}[!ht]
	\centering
	\includegraphics[width=.8\textwidth]{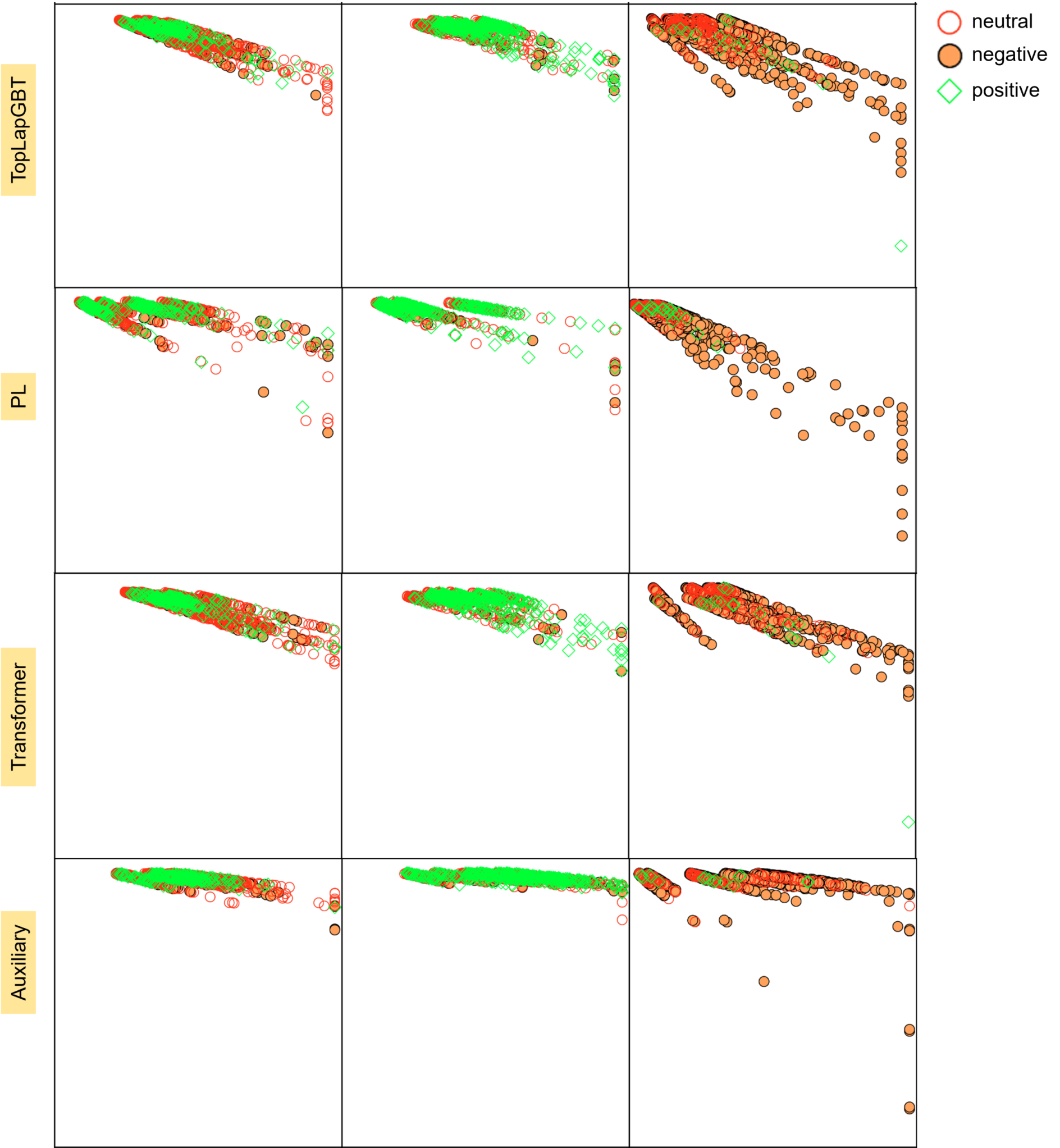}
	\caption{The comparison of R-S plots between the different types of features used in TopLapGBT model. The $y$-axis represents the residue score, whereas the $x$-axis represents the similarity score. RS scores were computed for the testing set, and all 10-folds were visualized. Each section corresponds to one of the 3 true solubility labels, and the sample's color and marker correspond to the predicted label from TopLapGBT.}
	\label{fig:rsplot}
\end{figure}

Figure \ref{fig:rsplot} juxtaposes the R-S plots derived from TopLapGBT against those from various feature sets utilized in model training. Across all visualizations, samples manifest a range of classification outcomes—both correct and incorrect—for each true label. However, a noteworthy observation is that Transformer-pretrain and persistent Laplacian-based features demonstrate superior clustering attributes compared to auxiliary features. The high RI and SI scores for auxiliary features cause these data points to cluster near the upper regions of their respective sections. Despite this, the integrative use of all three feature types in TopLapGBT results in appreciable clustering performance, corroborated by the CPR metrics obtained in 10-fold cross-validation.
To solidify the rationale behind adopting robust supervised classifiers like TopLapGBT, we contrast the R-S plots with UMAP visualizations (shown in Figure S2). It becomes evident that UMAP plots fail to form clusters that are as distinct as those observed in R-S plots, thereby reinforcing the need for a specialized approach to classify mutation samples effectively.

The impetus for utilizing structure-based features stems from the multifaceted relationship that exists among protein sequence, structure, and solubility. Factors such as hydrophobicity, charge distribution, and intermolecular interactions contribute to the complexity of protein solubility. Traditional prediction methods, which often rely on empirical rules or rudimentary descriptors, fall short in capturing this intricate molecular interplay. By employing advanced mathematical techniques like persistent Laplacian (PL) coupled with machine learning algorithms, we can decipher the complex patterns and relationships embedded within protein sequences and structures. Persistent Laplacian, in particular, provides a robust mathematical representation that captures both the topological and homotopic evolution of protein structures. Furthermore, machine learning models rooted in advanced mathematics offer several advantages for classifying changes in protein solubility. These models are well-suited for handling high-dimensional and complex data sets, such as those involving protein sequences and structures. They are also capable of learning non-linear relationships and capturing nuanced dependencies that are often overlooked by traditional linear models. Importantly, these advanced models can adeptly manage class-imbalanced datasets, which are commonly encountered in protein solubility studies.

\section{Conclusion}
 
In the multifaceted quest to understand mutation-induced solubility changes, various scientific domains including quantum mechanics, molecular mechanics, biochemistry, biophysics, and molecular biology have made significant contributions. Despite these concerted efforts, state-of-art models  have limitations, as evidenced by their normalized CPR value of 0.656 even after employing feature selection methods. Persistent homology (PH) has emerged as a powerful tool for capturing the complexity of biomolecular structures and has achieved noteworthy success in drug discovery applications. However, its inability to capture the nuances of homotopic shape evolution, crucial for delineating molecular interactions in proteins, presents a critical shortcoming.

Our study introduces TopLapGBT, a novel model that integrates persistent Laplacian (PL) features with pretrained transformer features, thereby bridging the gap in capturing both topology and homotopic shape evolution. This innovative fusion leads to significant advancements in classification performance. Specifically, TopLapGBT achieves normalized CPR and GC$^2$ scores of 0.688 and 0.361, respectively, marking improvements of 4.88\% and 15.71\% over the state-of-the-art PON-Sol2. These findings are further corroborated by an independent blind test, where TopLapGBT continues to outperform existing models.

In summary, our proposed TopLapGBT model not only achieves superior performance over existing state-of-the-art methods but also introduces a more nuanced approach for the classification of protein solubility changes upon mutation. These results underscore the transformative potential of integrating geometric and topological features with machine learning in advancing the field of molecular biology.

\section{Materials and Methods}
In this section, we endeavor to elucidate key mathematical and computational foundations that are instrumental for the work presented in this study. Specifically, we delve into spectral graph theory, simplicial complex, and persistent Laplacian methods, highlighting their significance in capturing topological and spectral properties essential for the characterization of proteins. Additionally, we discuss machine learning and deep learning paradigms, focusing on their role in processing, analyzing, and interpreting these complex features, especially within the confines of test datasets and validation settings.

\subsection{Persistent Laplacian characterization of proteins}
\paragraph{Simplicial complex}
A simplicial complex is made up of a set of simplices and generalises beyond graph networks at higher dimensions \cite{munkres2018elements,zomorodian2005topology,Edelsbrunner:2010,Mischaikow:2013}. Every simplex is a finite set of vertices which can be interpreted as the atoms in a protein structure. Essentially, simplices can be a point (0-simplex), an edge (1-simplex), a triangle (2-simplex), a tetrahedron (3-simplex), or in higher dimensions, a $p$-simplex. In other words, a $k$-simplex $\sigma^k = \{v_0, v_1,\cdots, v_k\}$ is the convex hull formed by $k+1$ affinely independent points $v_0, v_1, \cdots, v_k$ as follows,

\begin{equation*}
\sigma^k = \biggl\{\lambda_0v_0+\lambda_1v_1+\lambda_2v_2+\cdots + \lambda_kv_k|\sum_{i=0}^{k} \lambda_i = 1;\forall i, 0\leq \lambda_i \leq 1\biggr\}
\end{equation*}

A geometric simplicial complex $K$ is a finite set of geometric simplexes that satisfy two essential conditions. First, any face of a simplex from $K$ is also in $K$. Second,  the intersection of any two simplexes in $K$ is either empty or shares faces. Commonly used methods to construct simplicial complexes are \v{C}ech complex, Vietoris-Rips complex, Alpha complex, Clique complex, Cubic complex, and Morse complex \cite{munkres2018elements,zomorodian2005topology,Edelsbrunner:2010,Mischaikow:2013}.

\paragraph{Chain Group} A $k$-th chain group $C_k$ is a free Abelian group generated by oriented $k$-simplices $\sigma^k$. A boundary operator $\partial_k: C_k \rightarrow C_{k-1}$ defined on an oriented $k$-simplex $\sigma^k$ can be written as
\begin{equation*}
\partial_k\sigma^k = \sum_{i=0}^k (-1)^i[v_0,v_1,v_2,\cdots,\hat{v_i},\cdots, v_k],
\end{equation*}
where $[v_0,v_1,v_2,\cdots,\hat{v_i},\cdots, v_k]$ is an oriented $(k-1)$-simplex, which is constructed by the all the vertices except $v_i$, i.e., removing $v_i$ from the simplex. The boundary operator satisfies $\partial_{k-1}\partial_k = 0$.

The adjoint of $\partial_k$, which is
\[
\partial_{k}^*:C_{k-1}\to C_{k},
\]
satisfies the inner product relation $\langle \partial_k(f), g \rangle=\langle f, \partial_{k}^*(g) \rangle,$ for every $f \in C_{k}$, $g \in C_{k-1}$. This will be used in the combinatorial Laplacian.

\paragraph{Combinatorial Laplacian}
For the $k$-boundary operator $\partial_k: C_k \rightarrow C_{k-1}$ in $K$, define $\mathbf{B}_k$ to be an $m\times n$ matrix representation of the boundary operator under the standard bases $\{\sigma_i^k\}^n_{i=1}$ and $\{\sigma_j^{k-1}\}^m_{j=1}$ of $C_k$ and $C_{k-1}$. Similarly, the matrix representation of $\partial_{k}^*$ is the transpose matrix $\mathbf{B}_k^\top$, with respect to the same ordered bases of the boundary operator $\partial_k$. 

More specifically, let $m$ and $n$ be the number of $(k-1)$-simplices and $p$-simplices respectively in a simplicial complex $\mathcal{K}$.
The $m\times n$ boundary matrix $\mathbf{B}_k$ has entries defined as follows:
\begin{equation*}\label{eqn:boundary}
\mathbf{B}_k (i,j) = \left\{ \begin{array}{ll}
1, & \text{if } \sigma_i^{k-1} < \sigma_j^k, \sigma_i^{k-1} \sim \sigma_j^k.\\
-1, & \text{if } \sigma_i^{k-1} < \sigma_j^k, \sigma_i^{k-1} \nsim \sigma_j^k.\\
0, & \text{if } \sigma_i^{k-1} \nless \sigma_j^k.
\end{array} \right.
\end{equation*}
where $1\leq i\leq m$ and $1\leq j\leq n$. Here, $\sigma_i^{k-1} < \sigma_j^k$ represents the $i$-th $(k-1)$-simplex $\sigma_i^{k-1}$ is a face of $j$-th $k$-simplex  $\sigma_j^k$ and $\sigma_i^{k-1} \sim \sigma_j^k$ indicates the coefficient of $\sigma_i^{k-1}$ in $\partial_{k}(\sigma_j^k)$ is $1$. Likewise, $\sigma_i^{k-1} \nless \sigma_j^k$ means that $\sigma_i^{k-1}$ is not a face of $\sigma_j^k$ and $\sigma_i^{k-1} \nsim \sigma_j^k$ indicates that the coefficient of $\sigma_i^{k-1}$ in $\partial_{k}(\sigma_j^k)$ is $-1$.

Then the $k$-combinatorial Laplacian or the topological Laplacian is a linear operator $\Delta_k: C_k(K) \rightarrow C_k(K)$
\begin{equation}
	\Delta_k := \partial_{k+1}\partial_{k+1}^* + \partial_{k}^*\partial_{k}.
\end{equation}
The $k$-combinatorial Laplacian exhibits an $n\times n$ matrix representation $L_k$ and is given by 
\begin{equation}
	\mathbf{L}_k = \mathbf{B}_{k+1}\mathbf{B}_{k+1}^\top + \mathbf{B}_k^\top \mathbf{B}_k. 
\end{equation}
In the case $k=0$, then $\mathbf{L}_0 = \mathbf{B}_1\mathbf{B}_1^\top$ since $\partial_{0}$ is a zero map.

The number of rows in $\mathbf{B}_k$ represents the number of $(k-1)$-simplices in $K$ and the number of columns refers to the number of $k$-simplices in $K$. Furthermore, the upper $k$-combinatorial Laplacian matrix is $\mathbf{L}_k^U= \mathbf{B}_{k+1}\mathbf{B}_{k+1}^\top$ and the lower $k$-combinatorial Laplacian matrix is $\mathbf{L}_k^L = \mathbf{B}_k^\top\mathbf{B}_k$. Recall that since $\partial_0$ is a zero map, hence $\mathbf{L}_0(K) = \mathbf{B}_1\mathbf{B}_1^\top$ with $\mathbf{B}_0$ being a zero matrix and $K$ being an oriented simplicial complex of dimension 1. In fact, the $0$-combinatorial Laplacian matrix $\mathbf{L}_0(K)$ is actually the graph Laplacian in spectral graph theory.

The above graph Laplacian matrices can be explicitly described in terms of the simplex relations. More precisely, $\mathbf{L}_0$ can be described as
\[
\mathbf{L}_0(i,j) = \begin{cases}
d(\sigma_i^0), & \text{if } i=j\\
-1, & \text{if } i\neq j \text{ and } \sigma_i^0\frown \sigma_j^0\\
0, & \text{if } i\neq j \text{ and } \sigma_i^0\not\frown\sigma_j^0,
\end{cases}
\]
which is equivalent to the graph Laplacian. Furthermore, when $k>0$, $\mathbf{L}_k$ can be expressed as
\[
\mathbf{L}_k(i,j) = \begin{cases}
d(\sigma_i^k)+k+1, & \text{if } i=j\\
1, & \text{if } i\neq j, \sigma_i^k\not\frown\sigma_j^k, \sigma_i^k\smile \sigma_j^k \text{ and } \sigma_i^k\sim\sigma_j^k\\
-1, & \text{if } i\neq j, \sigma_i^k\not\frown\sigma_j^k, \sigma_i^k\smile \sigma_j^k \text{ and } \sigma_i^k\not\sim\sigma_j^k\\
0, & \text{if } i\neq j, \text{ and either }\sigma_i^k\frown\sigma_j^k \text{ or }\sigma_i^k\not\smile \sigma_j^k.
\end{cases}
\]
Here, we denote $\sigma_j^{k-1} \sim \sigma_i^k$ if they have the same orientation, i.e. similarly oriented. Furthermore, we say that two $k$-simplices $\sigma_i^k$ and $\sigma_j^k$ are upper adjacent (resp. lower adjacent) neighbors, denoted as $\sigma_i^k \frown \sigma_j^k$ (resp. $\sigma_i^k \smile \sigma_j^k$), if they are both faces of a common $(k+1)$-simplex (resp. they both share a common $(k-1)$-simplex as their face). In addition, if the orientations of their common lower simplex are the same, it is called similar common lower simplex ($\sigma_i^k \smile \sigma_j^k$ and $\sigma_i^k \sim \sigma_j^k$). On the other hand, if the orientations are different, it is called dissimilar common lower simplex ($\sigma_i^k \smile \sigma_j^k$ and $\sigma_i^k \nsim \sigma_j^k$). The (upper) degree of a $k$-simplex $\sigma_i^k$, denoted as $d(\sigma_i^k)$, is the number of $(k+1)$-simplices, of which $\sigma_i^k$ is a face.

The eigenvalues of combinatorial Laplacian matrices are independent of the choice of the orientation \cite{horak2013spectra}. Furthermore, the multiplicity of zero eigenvalues, i.e. the total number of zero eigenvalues, of $\mathbf{L}_k$ corresponds to the $k$th Betti number $\beta_k$, according to the combinatorial Hodge theorem \cite{eckmann1944harmonische}. The $k$th Betti numbers are topological invariants that describe the $k$-dimensional holes in a simplicial complex. In particular, $\beta_0$, $\beta_1$ and $\beta_2$ represents the numbers of independent components, rings and cavities, respectively.
\paragraph{Persistent Laplacian}

Persistent Laplacian (PL) were first introduced by integrating graph Laplacian and multiscale filtration \cite{wang2020persistent}. Analyzing the spectra of $k$-combinatorial Laplacian matrix allows both topological and geometric information (i.e. connectivity and robustness of simple graphs) to be obtained. However, this method is genuinely free of metrics or coordinates, which induced too little topological and geometric information that can be used to describe a single configuration. 

Therefore, PL was extended to simplicial complexes. This allows a sequence of simplicial complexes from a filtration process to generate persistent Laplacian which is largely inspired by persistent homology and in earlier works in multiscale graphs. For the rest of this section, we introduce mainly on the  construction of PL. First, a $k$-combinatorial Laplacian matrix is symmetric and positive semi-definite. Therefore, its eigenvalues are all real and non-negative. The multiplicity of zero spectra (also called harmonic spectra) reveals
the topological information, and the geometric information will be preserved in the non-harmonic spectra.

A key concept of PL is the filtration process. Essentially, an ever-increasing filtration value $f$ is used to generate a series of topological spaces, which are represented by a nested sequence of multiscale simplicial complexes.  Naturally, PL generates a sequence of simplicial complexes induced by varying a filtration parameter \cite{wang2020persistent}. For an oriented simplicial complex $K$, its filtration is a nested sequence of simplicial complexes
$(K_t)^m_{t=0}$ of $K$ 
\begin{equation*}
\varnothing = K_0 \subseteq K_1 \subseteq \cdots \subseteq K_m=K.
\end{equation*}

This nested sequence of simplicial complexes induces a family of chain complexes 
\begin{equation}
	\left\{\cdots \underset{\partial_{k+2}^{t*}}{\stackrel{\partial_{k+2}^{t}}{\leftrightharpoons}} C^t_{k+1} \underset{\partial_{k+1}^{t*}}{\stackrel{\partial_{k+1}^{t}}{\leftrightharpoons}} C^t_k \underset{\partial_{k}^{t*}}{\stackrel{\partial_{k}^{t}}{\leftrightharpoons}} \cdots \underset{\partial_{1}^{t*}}{\stackrel{\partial_{1}^{t}}{\leftrightharpoons}} C^t_0 \underset{\partial_{0}^{t*}}{\stackrel{\partial_{0}^{t}}{\leftrightharpoons}} \varnothing\right\}_{t\in \R^+}.
\end{equation}
where $C^t_k=C_k(K_t)$ is the chain group for the subcomplex $K_t$, and its $k$-boundary operator is $\partial^t_k:C_k(K_t)\rightarrow C_{k-1}(K_t)$. In the case $k<0$, then $C_k(K_t)$ is an empty set and $\partial_k^t$ is a zero map. For $0<k<\dim(K_t)$, the boundary operator 
\begin{equation}
	\partial_{k}^t(\sigma_k) = \sum_{i=0}^k (-1)^i\sigma^{k-1}_i, \quad \sigma^k\in K_t, 
\end{equation}
with $\sigma^k = [v_0, \cdots, v_k]$ being the $k$-simplex, and $\sigma^{k-1}_i = [v_0,\cdots,\hat{v_i}, \cdots, v_k]$ being the $(k-1)$-simplex for which its vertex $v_i$ is removed. Similarly, the adjoint of $\partial_{k}^t$ is the operator $\partial_{k}^{t*}: C_{k-1}(K_t) \rightarrow C_k(K_t)$. The topological and spectral characteristics can then be studied from  $\mathbf{L}_k(K_t)$ by varying the filtration parameter and diagonalizing the $k$-combinatorial Laplacian matrix. The multiplicity of the zero spectra of $\mathbf{L}_k^t$ is the persistent Betti number $\beta_k^t$, which represents the number of $k$-dimensional holes in $K_t$. In other words, 
\begin{equation}
	\beta_k^t = \dim(L_k^t) - \text{rank}(\mathbf{L}_k^t) = \text{nullity}(\mathbf{L}_k^t)=\# \text{ of harmonic spectra of }\mathbf{L}_k^t.
\end{equation}

In particular, $\beta_0^t$ represents the number of connected components in $K_t$, $\beta_1^t$ counts the number of one-dimensional cycles in $K_t$ and $\beta_2^t$ reveals the number of two-dimensional voids in $K_t$. In addition, the spectra of $\mathbf{L}_k^t$ can be written in the following ascending order
\begin{equation}
	\text{Spectra}(\mathbf{L}_k^t) = \{(\lambda_1)_k^t, (\lambda_2)_k^t, \cdots, (\lambda_n)_k^t\},
\end{equation}
where $\mathbf{L}_k^t$ here is an $n\times n$ matrix. The $p$-persistent $k$-combinatorial Laplacian can be extended based on the boundary operator as well. Further details can be found in \cite{wang2020persistent}.

\begin{figure}[!ht]
	\centering
	\includegraphics[width=.8\textwidth]{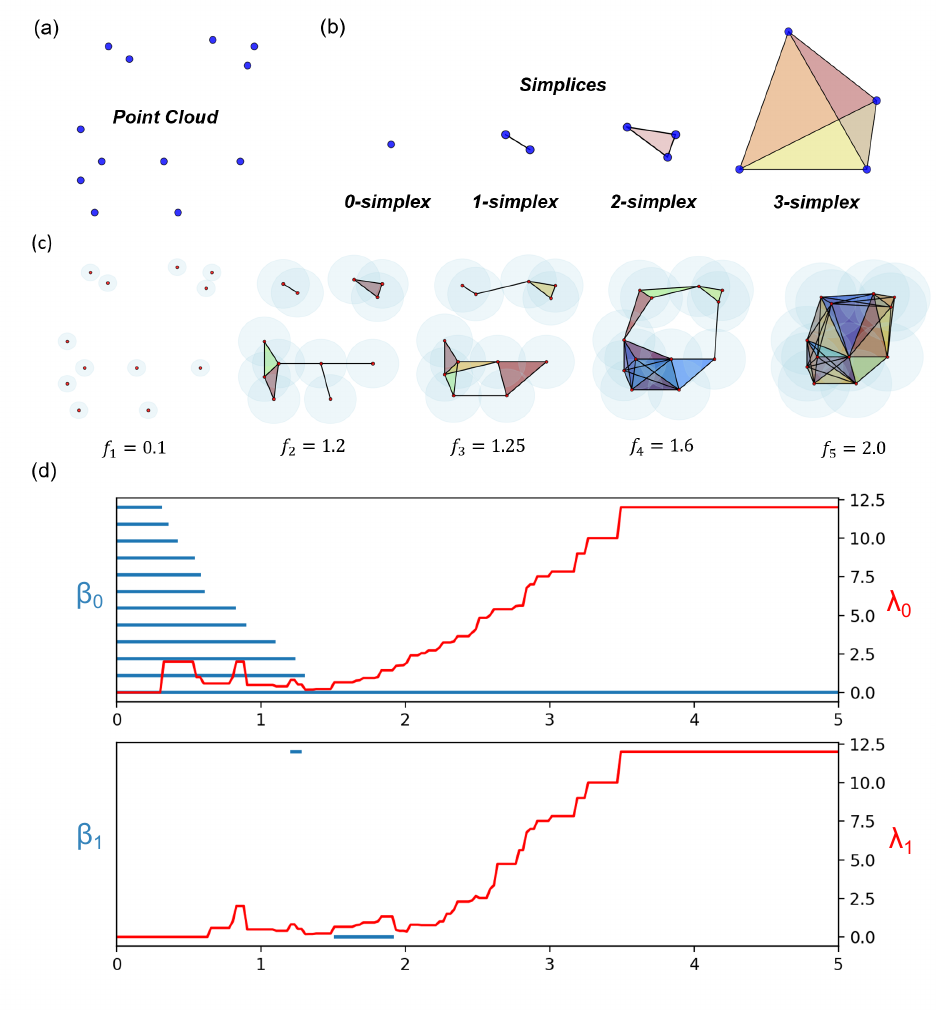}
	\caption{The illustration of (a) point cloud, (b) basic simplices and (c) the filtration process and (d) Comparison between PH barcodes \cite{zomorodian2012topological,edelsbrunner2022computational} and the non-harmonic spectra of persistent Laplacians (PLs) \cite{wang2020persistent} from the filtration process in (c). The $x$-axis represents the filtration parameter $f$. By discretising the filtration region into equal-sized bins and adding all the Betti bars together, the topological invariants are summarized into persistent Betti numbers that acts a topological descriptor extracted from protein structures. Persistent Laplacians (PLs) \cite{wang2020persistent} for thirteen points. The first non-zero eigenvalues of dimension 0, $\lambda_0(r)$, and dimension 1, $\lambda_1(r)$, of PLs are depicted in red. The harmonic spectra of PLs return all the topological invariants of PH, whereas the non-harmonic spectra of PLs capture the additional homotopic shape evolution of PLs during the filtration that are neglected by PH.
	}
	\label{fig:TDA}
\end{figure}

In order to illustrate the difference between PL and PH, Figure \ref{fig:TDA} describes a point cloud, basic simplices, a filtration process and the comparison between persistent Laplacian and persistent homology barcodes of 13 points. The filtration process in Figure \ref{fig:TDA}(c) shows the different stages of a Rips filtration process for the 13 points. Figure \ref{fig:TDA}(d) shows the
persistent homology barcodes (in blue) and persistent non-harmonic spectra (in red). It can be seen that the non-harmonic spectra provides the additional  homotopic shape evolution that is missing in persistent homology in the later part of the filtration process.

\subsection{Persistent Laplacian descriptors}
\begin{figure}[!ht]
	\centering
	\includegraphics[width=.9\textwidth]{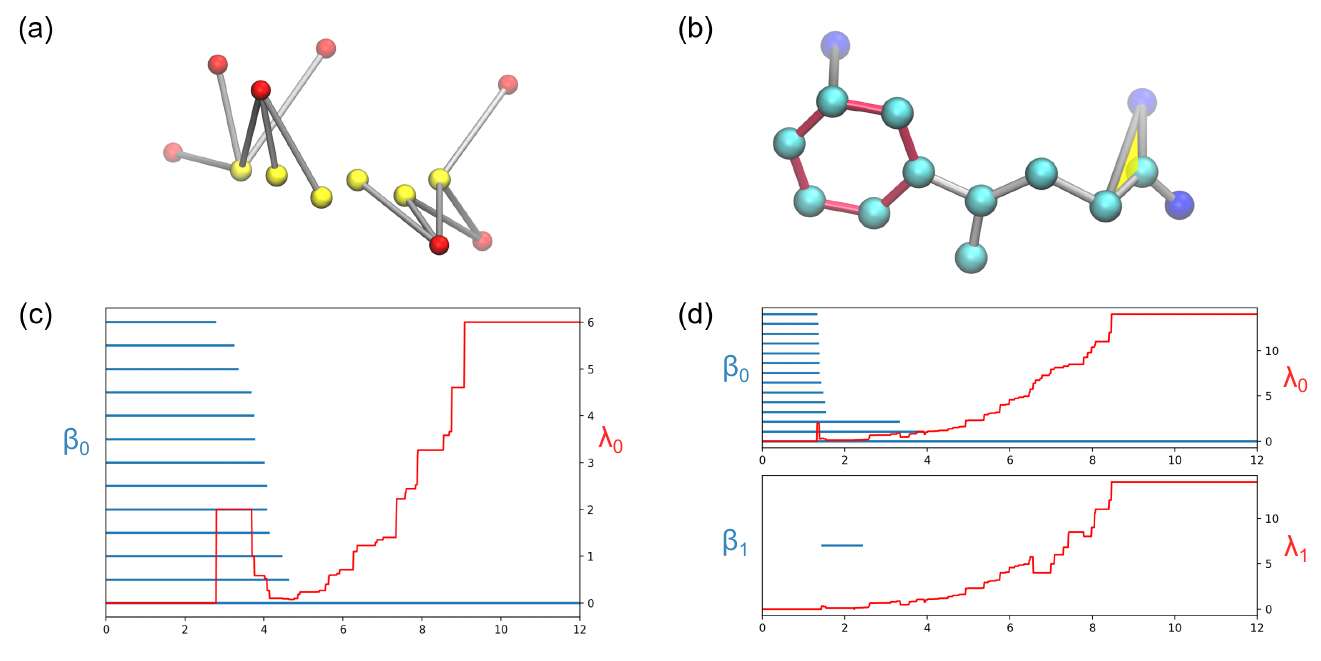}
	\caption{An illustration of interactive PL showing hydrophillic interactions based on DI-based filtration (left) and hydrophobic interactions based on DE-based filtration (right) at a mutation site. (a) Hydrophillic interactions between the nitrogen atoms (red) and oxygen atoms (yellow). (b) Hydrophobic interactions between the carbon atoms (cyan) and oxygen atoms (dark blue). The hexagon ring is colored in red and the triangle is colored in yellow. (c) The PH barcodes and PL for dimension 0 of the hydrophillic interactions in (a). (d) The PH barcodes and PL for dimension 0 and dimension 1 of the hydrophobic interactions in (b). The Betti-1 bar is due to the red hexagon ring in (b).
	}
	\label{fig:interaction}
\end{figure}

In order to capture the mutation-induced solubility change, we apply the persistent Laplacian (PL) to characterize the interactions between the mutation site and the rest of the protein. To describe these interactions, we first propose the interactive PL with the distance function ${\rm DI}(A_i,A_j)$ describing the distance between two atoms $A_i$ and $A_j$ defined as 
\begin{equation}
	{\rm DI}(A_i, A_j) = \begin{cases}
		\infty, & \text{if Loc}(A_i) = \text{Loc}(A_j),\\
		{\rm DE}(A_i, A_j), & \text{otherwise}.
		\end{cases}
\end{equation}
where ${\rm DE}(\cdot, \cdot)$ is the Euclidean distance between the two atoms and $\text{Loc}(\cdot)$ refers to the atom's location which is either in the mutation site or in the rest of the protein. Here, we construct two types of simplicial complexes in our PL computation, such as Vietoris-Rips complex (VC) and Alpha complex (AC). Both complexes are used to characterize the first order interactions and higher order patterns respectively. To capture and characterize different types of atom-atom interactions, we generate the PL based on different atom subsets by selecting one type of atom in the mutation site and one other atom type in the rest of the protein. Different types of atom-atom interactions characterize the different interactions in proteins. For example, interactions generated from carbon atoms are associated with hydrophobic interactions. Similarly, interactions between nitrogen and oxygen atoms correlate to hydrophilic interactions and/or hydrogen bonds. Both types of interactions are illustrated in Figure \ref{fig:interaction}. Interactive PLs have the capability to unveil additional details about bonding interactions and offer a fresh and distinct representation of molecular interactions in proteins.

The set of persistent spectra from each persistent Laplacian computation 
consists of $V^{DI}_{\gamma, \alpha, \beta}$ and $V^{DE}_{\gamma, \alpha, \beta}$ where $\gamma \in \{M, W\}$ refers to the mutant protein or the wild type protein, $\alpha \in \{C,N,O\}$ is the atom type chosen in the rest of the protein and $\beta \in \{C,N,O\}$ is the atom type chosen in the mutation site. $V^{DI}_{\gamma, \alpha, \beta}$ applies the distance DI-based filtration to generate 0-dimensional Laplacian using the Vietoris-Rips complex and $V^{DE}_{\gamma, \alpha, \beta}$ applies the Euclidean distance DE-based filtration to generate 1 and 2-dimensional Laplacian using the  alpha complex. In total, there are 54 sets of persistent spectra. The persistent spectra from PL contains both harmonic and non-harmonic spectra that are capable of revealing the molecular mechanism of protein solubility.

For zero dimensions, we consider both the harmonic spectra and non-harmonic spectra information for each persistent Laplacian. Filtration using Rips complex with $DI$ distance is used. The 0-dimensional PL features are generated from 0\AA\space to 6\AA\space with 0.5\AA\space gridsize.
For the non-harmonic spectra information, we count the number of non-harmonic spectra and calculate seven statistical values of non-harmonic spectra such as sum, minimum, maximum, mean, standard deviation, variance and the sum of eigenvalues squared. This generates eight statistical values for each of the nine atomic pairs. Therefore, the dimension of 0-dimensional PL features for a protein is 72. In total, the 0-dimensional PL-based feature size after concatenating features at different dimensions for wild type and mutant is 1872.

For one or two dimensions, we perform the filtration using Alpha complex with the $DE$ distance. The limited number of atoms in the local protein structure can create only a few high-dimensional simplexes, resulting in minimal alterations in shape. As a result, it suffice to consider features from only harmonic spectra of persistent Laplacians by coding the topological invariants for the high-dimensional interactions. Using GUDHI\cite{gudhi:urm}, the persistence of the harmonic spectra can be represented by persistent barcodes. The topological feature
vectors are generated by computing the statistics of bar lengths, births and
deaths. Bars shorter than 0.1\AA\space are excluded as they do not exhibit
any clear physical meaning. The remaining bars are then used for computing the statistics: (1) sum, maximum and mean for lengths of bars; (2) minimum and maximum for the birth values of bars; (3) minimum and maximum for the death values of bars. Each set of point clouds leads to a seven-dimensional vector. These features
are calculated on nine single atomic pairs and one heavy atom pair. The dimension of one- and two-dimensional PL feature vectors for a protein is 140. In total, the higher-dimensional PL-based feature size after concatenating features at different dimensions for wild type, mutant and their difference is 420.

\subsection{Persistent Homology}
Persistent homology is part of the harmonic spectra of PL. The homology groups in PH illustrate the persistence of topological invariants, hence providing the harmonic spectral information in PL. The site- and element-specific PH features are generated in a similar way as compared to PL. Similar to PL, filtration construction is also employed to PH. For the zero dimension, the filtration parameter can be discretized into several equally spaced bins, namely [0, 0.5], (0.5, 1], $\cdots$, (5.5, 6]\AA. The death value of the bars are summed in each bin resulting in 12$\times$18 features. 

For each bin, we count the numbers of persistent bars, resulting in a
nine-dimensional vector for each point cloud. Similarly, this is performed for each of the nine single atomic pairs. Hence, the dimension of PH features for a protein is 216. For one or two dimensions, the identical featurization from the statistics of persistent bars in PH is used. The PH embedding combines features at different dimensions as described above and concatenated for wild type, mutant and their difference, resulting in a 648-dimensional vector.

\subsection{Transformer Features}
Recently, we have seen significant advancements in modelling protein properties using large-scale protein transformer models trained on hundreds of millions of sequences. These models, like ESM \cite{rives2021biological} (evolutionary scale modeling) and ProtTrans\cite{vaswani2017attention, Devlin2019BERTPO}, have demonstrated impressive performance. Moreover, hybrid fine-tuning approaches that leverage both local and global evolutionary data have proven to enhance these models even further. For instance, eUniRep is an improved LSTM-based UniRep model achieved through fine-tuning with knowledge extracted from local multiple sequence alignments (MSAs). Additionally, the ESM model can be fine-tuned using either downstream task data or local MSAs. In our research, we employed the ESM-1b transformer, a model that falls under the transformer architecture. This particular variant was trained on a dataset of 250 million sequences using a masked filling procedure and boasts an architecture comprising 34 layers with a whopping 650 million parameters. The ESM transformer's primary role in our work was to generate sequence embeddings. At each layer of the ESM model, it encoded a sequence of length L into a matrix sized at 1,280$\times$L, excluding the start and terminal tokens. For our study, we utilized the sequence representation derived from the final (34th) layer and computed the average along the sequence length axis, resulting in a 1,280-component vector.

\subsection{Performance Metrics}
PPV and NPV assesses the true positive and true negative proportion of the predicted results for each solubility class. PPV and NPV are computed based on TP, TN, FP and FN which represents the true positive, true negative, false positive and false negative values for each solubility class. For each solubility class, PPV and NPV can be computed by:
\begin{equation}
	\rm PPV = \frac{TP}{TP+FP}.
\end{equation}
\begin{equation}
	\rm NPV = \frac{TN}{TN+FN}.
\end{equation}
Furthermore, specificity and sensitivity can be computed by the following:
\begin{equation}
	\rm Specificity = \frac{TN}{TN+FP}.
\end{equation}
\begin{equation}
\rm Sensitivity = \frac{TP}{TP+FN}.
\end{equation}
The correct prediction ratio (CPR) and generalized squared correlation (GC$^2$) are used to evaulate the overall performance of TopLapGBT. CPR and GC$^2$ can be computed as 
\begin{equation}
	{\rm CPR} = \frac{1}{N}\sum_{i} z_{ii}, \text{ and}
\end{equation}
\begin{equation}
	{\rm GC^2} = \frac{1}{N(K-1)}\sum_{ij} \frac{(z_{ij}-e_{ij})^2}{e_{ij}},
\end{equation}
where $K$ is the number of classes and $N$ is the number of samples. Here, $z_{ij}$ represents the number of samples of class $i$ to class $j$. Let $x_i=\sum_{j}z_{ij}$ be the number of inputs from class $i$, and $y_j = \sum_{i}z_{ij}$ be the number of inputs predicted to class $j$. Then the expected number of samples in $(i,j)$-th entry of the multiclass confusion matrix is 
\[
e_{ij} = \frac{x_iy_j}{N}.
\]
Since the mutational samples across the three solubility classes are imbalanced, we normalized the values to provide more reliable calculation of performance metrics.

\section{Software and resources}
Protein sequences are first preprocessed by AlphaFold 2 to generate wild type protein structures. In particular, 3D protein structures are generated from protein sequences using ColabFold \cite{mirdita2022colabfold}. Mutant proteins are generated from the Jackal software\cite{xiang2002jackal}.
All TopLapGBT models are built using the sklearn machine learning library \cite{scikit-learn}. The hyperparameters for all the TopLapGBT are: n\_estimators = 20000, learning\_rate = 0.05, max\_depth = 7, subsample=0.4, min\_sample\_split = 3 and max\_features = sqrt. The PQR files, which contains the partial charge information of the proteins, are generated from the PDB2PQR software \cite{dolinsky2004pdb2pqr}. The PQR files for both the wild type proteins are generated with AMBER force field. The solvation energy and surface area information are calculated from the in-house online software package ESES \cite{liu2017eses} and MIBPB \cite{chen2011mibpb}. The pKa values are computed from the PROPKA software package \cite{li2005very}. The position-specific-scoring matrices (PSSM) are computed from the BLAST+ software \cite{johnson2008ncbi} using the nr database. The secondary structure features and torsion angle sequence-based information are calculated from SPIDER \cite{heffernan2015improving}. The persistent Laplacian descriptors for both VR complexes and alpha complexes are calculated using the GUDHI software library \cite{maria2014gudhi}. All computational work in support of this research was performed using the resources from the National Super Computing Centre of Singapore (NSCC).

%
%

\section*{Code and Data Availability}
The 3D protein structures and the TopLapGBT code can be found in \linebreak
\href{https://github.com/ExpectozJJ/TopLapGBT}{https://github.com/ExpectozJJ/TopLapGBT}. The source code for the R-S plot can be found at \href{https://github.com/hozumiyu/RSI}{https://github.com/hozumiyu/RSI}.

\section*{Supporting Information}
Supporting Information is available for supplementary tables, figures, and methods. 

\section*{Acknowledgments}
This work was supported in part by NIH grants  R01GM126189, R01AI164266, and R01AI146210, NSF grants DMS-2052983,  DMS-1761320, and IIS-1900473,  NASA grant 80NSSC21M0023,  MSU Foundation,  Bristol-Myers Squibb 65109, and Pfizer.
It was supported in part by Nanyang Technological University Startup Grant M4081842.110, Singapore Ministry of Education Academic Research fund Tier 1 RG109/19 and Tier 2 MOE-T2EP20120-0013, MOE-T2EP20220-0010, and MOE-T2EP20221-0003.  

\vspace{0.6cm}
\bibliographystyle{ieeetr}
\bibliography{refs}



\end{document}